\documentclass[conference]{IEEEtran}

\usepackage{graphicx}

\begin{document}
%
%
\title{Do retinal ganglion cells project natural scenes to their principal subspace and whiten them?}


%
\author{\IEEEauthorblockN{Reza Abbasi-Asl\IEEEauthorrefmark{1}\IEEEauthorrefmark{2},
Cengiz Pehlevan\IEEEauthorrefmark{2},
Bin Yu\IEEEauthorrefmark{1}\IEEEauthorrefmark{3}, 
and Dmitri Chklovskii\IEEEauthorrefmark{2}}
\IEEEauthorblockA{\IEEEauthorrefmark{1}Department of Electrical Engineering and Computer Science,  University of California,  Berkeley, CA.}
\IEEEauthorblockA{\IEEEauthorrefmark{2} Center for Computational Biology, Flatiron Institute, New York, NY.}
\IEEEauthorblockA{\IEEEauthorrefmark{3} Department of Statistics, University of California, Berkeley, CA.}}


\maketitle

\begin{abstract}


Several theories of early sensory processing suggest that it whitens sensory stimuli. Here, we test three key predictions of the whitening theory using recordings from 152 ganglion cells in salamander retina responding to natural movies. We confirm the previous finding that firing rates of ganglion cells are less correlated compared to natural scenes, although significant correlations remain. We show that while the power spectrum of ganglion cells decays less steeply than that of natural scenes, it is not completely flattened. Finally, we find evidence that only the top principal components of the visual stimulus are transmitted.
\end{abstract}


%
\IEEEpeerreviewmaketitle

\section{Introduction}

Efficient coding theories have been extremely influential in understanding early sensory processing in general \cite{attneave1954some,barlow1961possible} and, visual processing in the retina in particular \cite{srinivasan1982predictive,huang2011predictive,atick1992does,atick1992could,van1992theory,haft1998theory,doi2012efficient}. 
Many efficient coding theories predict that the retina should whiten sensory stimuli \cite{graham2006can}. For instance, whitening is optimal from the predictive coding theory perspective, which postulates that only the unpredicted part of a signal should be transmitted. An optimal linear predictive coding filter is a whitening filter, and when applied to the retina this theory predicts that the receptive fields of retinal ganglion cells whiten visual stimuli \cite{srinivasan1982predictive,huang2011predictive}. Redundancy reduction theory proposes that redundant information transmission at the retinal output should be minimized by decorrelated firing of ganglion cells \cite{atick1992does}. Information theory, too, favors whitening. A linear filter that maximizes mutual information between noisy inputs and noisy outputs is a whitening filter \cite{van1992theory,haft1998theory,linsker1988self}. Finally, whitening was proposed by another normative theory \cite{pehlevan2015normative}, whose starting point was efficient dimensionality reduction through similarity matching \cite{pehlevan2015hebbian}.

 

Here, using electrophysiological recordings from 152 ganglion cells in salamander retina responding to natural movies, we test three key predictions of the whitening theory. 1) Retinal ganglion cells should be uncorrelated. We confirm previous tests of this prediction finding that retinal ganglion cells are indeed decorrelated compared to strong spatial correlations of natural scenes \cite{pitkow2012decorrelation}, but significant redundancy remains \cite{puchalla2005redundancy}. 2) Output power spectrum should be flat. Whereas we observe flattening of the ouput compared to the input, it is not completely flat. 3) If the number of channels are reduced from the input to the output, the top input principal components are transmitted. 

We start by detailing the dataset we used for this work. Then, we present our results, which test the three predictions. Our results show that perfect whitening is not the whole story, consistent with findings of \cite{puchalla2005redundancy,pitkow2012decorrelation}.

\section{Experimental Dataset and Notation}

The dataset was collected by Michael Berry's laboratory at Princeton University. It consists of recorded responses from 152 larval tiger salamander retinal ganglion cells to a 7-minute natural movie stimulus. The stimulus consists of gray scale movie of leaves and branches blowing in the wind (figure \ref{fig:decorr}-A) and is projected onto the array from a CRT monitor at 60 Hz. The frame size of the stimulus is $512 \times 512$. Details of recording procedure, including the spike-sorting algorithm, are described in \cite{prentice2015error}.

In this paper, $\bf X$ and $\bf Y$ denote the stimulus and the firing rate matrices respectively. $\bf X$ is a $n\times T$ matrix, where $n$ is the total number of pixels in each frame (reshaped to form a vector) and $T$ is the number of time samples. In our setting, for computational purposes, stimulus frames are down-sampled to $n=128\times128 = 16,384$ and the 7-minute movie shown at 60 Hz corresponds to $T=25,200$. Moreover, $\bf Y$ is the $k\times T$ firing rate matrix, where $k=152$ is the number of ganglion cells that data is recorded from. The firing rates are reported based on spike counts in sliding 80 ms time windows. Both $\bf X$ and $\bf Y$ are centered in time.


%

\section{Results}

\subsection{Do retinal ganglion cells decorrelate natural scenes?}

\begin{figure}
     \centering
     \includegraphics[scale=.31,clip,trim=0in 1.2in 0in 0in]{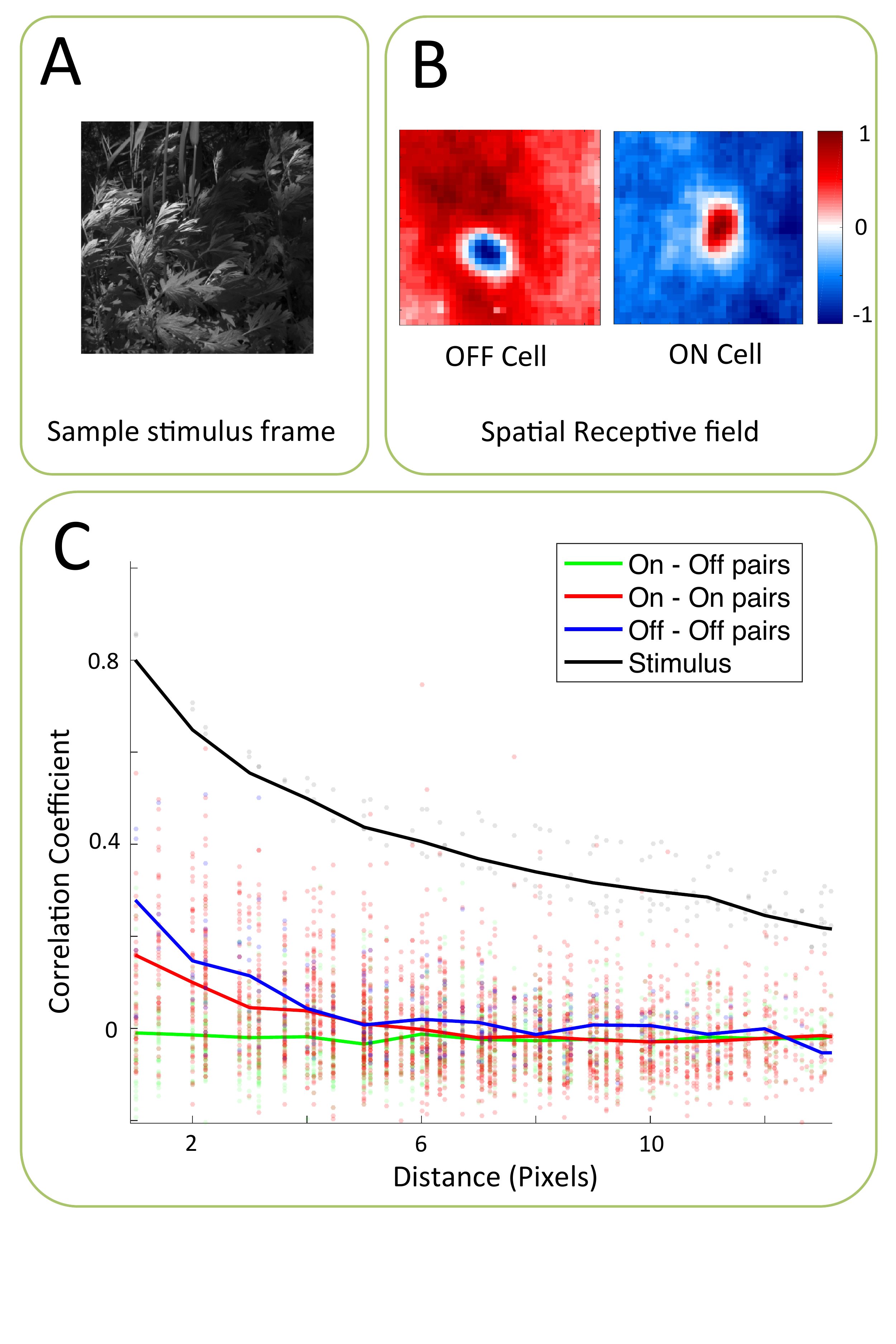}
     \caption{Output of ganglion cells are decorrelated compared to natural scenes. A. A sample frame from stimulus. B. Estimated receptive field for an On and an Off ganglion cell. C. Pearson correlation coefficient of firing rates for all ganglion cell-type pairs as a function of the distance between their receptive field centers. The curves corresponds to median correlation coefficient in a distance window of 1 pixel. Similar correlation coefficient and curve are shown for stimulus.} 
     

     \label{fig:decorr}
 \end{figure}

We start by testing whether output of ganglion cells are indeed decorrelated. This question was to a large extent answered by previous studies \cite{puchalla2005redundancy,pitkow2012decorrelation}, and here we confirm their findings. 


Our goal is to study the correlation between firing rates of ganglion cells as a function of distance between their receptive field center, therefore, it is necessary to first estimate the receptive field of each cell. We use regularized spike-triggered average also known as regularized reverse correlation technique \cite{chichilnisky2001simple} to estimate the spatial receptive field of each cell. In this technique, firing rate matrix is assumed to be a linear function of stimulus matrix, $\bf Y = RX $. Here, $\bf R$ is the $k \times n$ receptive field matrix for ganglion cells. $\bf R$ can be estimated using ridge regression which has the following closed form:
$$\bf R = YX^T(XX^T+\alpha I)^{-1}.$$ 	
Here, $\alpha$ is regularization parameter which we chose with 10-fold cross validation and $\bf I$ is identity matrix. Examples of the estimated On and Off receptive fields are shown in figure  \ref{fig:decorr}-B. The center of receptive field for each cell is simply identified as the location with maximum (for On cells) or minimum (for Off cells) value of estimated receptive field. 

Figure \ref{fig:decorr}-C shows the Pearson correlation coefficient (CC) of firing rates for all ganglion cell-type pairs as a function of the distance between their receptive field centers. This is depicted for On-On, Off-Off, and On-Off pairs.  To show the trend of each group, the median correlation coefficient in a distance window of 1 pixel is also shown. We compare this correlation with the spatial correlation in the stimulus. Consistent with Pitkow and Meister's results \cite{pitkow2012decorrelation}, firing rates of ganglion cells are less correlated than the pixels in naturalistic stimulus. However, significant correlations remain, questioning the validity of the whitening theory (see also \cite{puchalla2005redundancy}).

\subsection{Is the power spectrum of ganglion cells flattened?}

\begin{figure*}[t!]
     \centering
     \includegraphics[scale=.3,clip,trim=0in 3.8in 0in 0in]{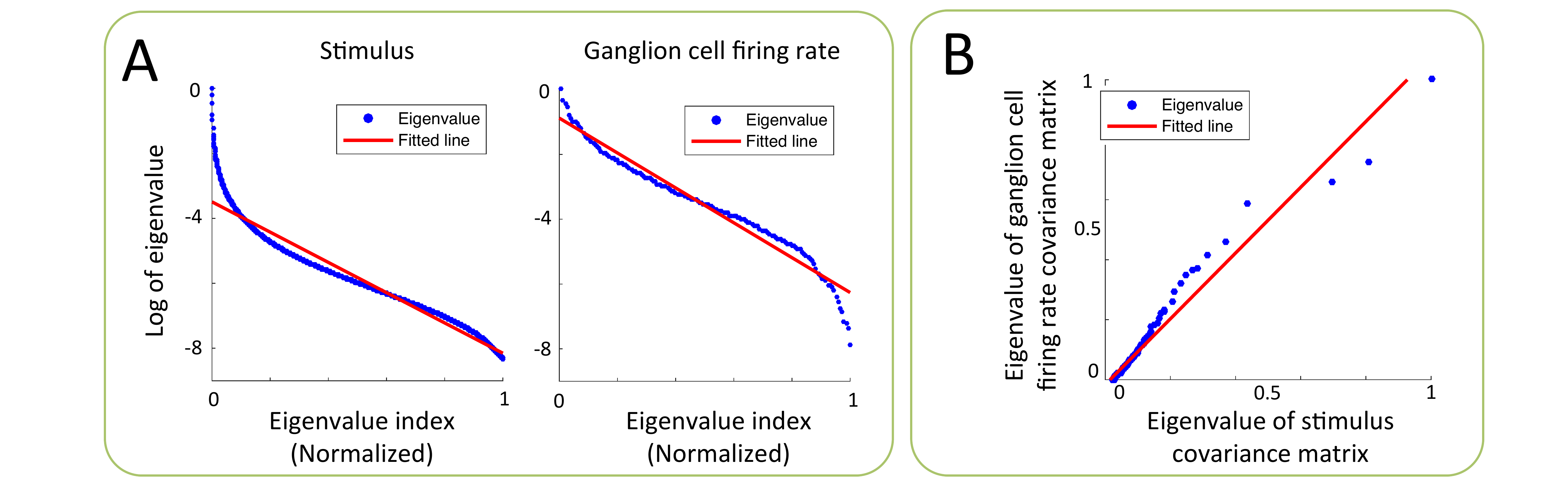}
     \caption{Power spectrum of ganglion cells are not flattened. A. Eigenvalues of stimulus and ganglion cells firing rate covariance matrices. Eigenvalues are normalized by the largest eigenvalue and are shown in log scale. The index is also normalized by the largest value. A line can be fitted on each plot with slope -4.7 and -5.3, respectively. B. Scatter plot of top 152 eigenvalue pairs. A line can be fitted to this plot with slope 1.09. }

     \label{fig:psd_decay}
\end{figure*}
 
Next we test whether the power spectrum of ganglion cells is flattened. In Figure \ref{fig:psd_decay}-A, we show the eigenvalues of the ganglion cell firing rate covariance matrix, normalized by the highest eigenvalue. We also show similar plot for the visual stimuli covariance matrix for comparison. We observe a decay in the power spectrum of the firing rates. 

To compare the rate of power spectrum decay, we plot the sorted eigenvalues of the firing rate covariance matrix against top 152 eigenvalues of natural stimuli covariance matrix, Figure \ref{fig:psd_decay}-B, and observe that a line can be fitted to this plot with slope 1.09, suggesting again that the decay is similar (A perfectly equalized output spectrum would lead to a slope of 0). This analysis assumed that in the scatter plot of Figure \ref{fig:psd_decay}-B, the correct correspondence between stimulus and firing-rate covariance eigenvalues is in order of their magnitude. It is possible that due to noise, we misestimate the rank-order of eigenvalues. We explore this possibility in more detail below. 


%

\subsection{Do retinal ganglion cells project natural scenes to their principal subspace?}

Finally we test whether retinal ganglion cell outputs represent a projection of natural scenes to their principal subspace. To be mathematically precise, we want to test if ${\bf Y} = {\bf O}{\bf D}{\bf P}{\bf X}$, where {\bf P} is a $k\times n$ matrix with $k<n$, rows of ${\bf P}$ are the principal eigenvectors of the stimulus covariance matrix, ${\bf D}$ is a diagonal, nonnegative, $k\times k$ matrix, which could be rescaling power in each component for e.g. whitening, and ${\bf O}$ is an arbitrary $k$-dimensional orthogonal matrix. Since we do not know ${\bf O}{\bf D}$, we do not have a direct prediction for what ${\bf Y}$ should be. 

However, a prediction can be made for the right-singular vectors of ${\bf Y}$. Suppose a singular-value decomposition (SVD) for ${\bf X}={\bf U}_{X}{\bf \Lambda}_{X}{\bf V}_{X}^T$, where singular values are ordered in decreasing order. Then,  ${\bf Y} = {\bf O}{\bf D}{\bf P}{\bf U}_{X}{\bf \Lambda}_{X}{\bf V}_{X}^T$. But, rows of ${\bf P}$ are the first $k$ columns of ${\bf U}_X$, and hence ${\bf \Lambda}_{Y} := {\bf D}{\bf P}{\bf U}_{X}{\bf \Lambda}_{X}$ is a $k\times N$ diagonal matrix. Then, ${\bf Y}={\bf R}{\bf \Lambda}_{Y}{\bf V}_{X}^T$ is an SVD, and we can claim that the top $k$ right singular vectors of ${\bf X}$ and ${\bf Y}$ (columns of ${\bf V}_{ X}$) should match. 

Of course the match will not be perfect due to noise, which may effect our analysis in two ways. First, our estimate of the ordering of singular vectors will be imperfect. Second, the singular vector estimates will be imperfect. 

 \begin{figure*}[t!]
      \centering
      \includegraphics[scale=.2,clip,trim=.7in .5in .7in .1in]{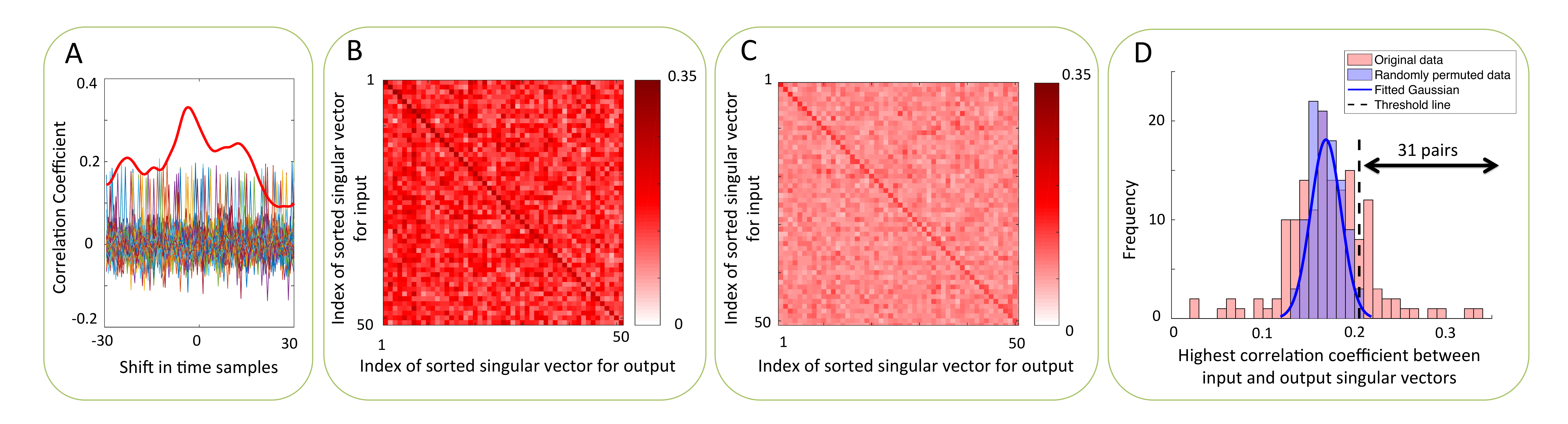}
      \caption{Input-output singular vector pair identification. A. Red curve shows correlation coefficient between the input-output right singular vector pair with the highest peak value among other pairs. Other curves show correlation coefficients between right singular vectors for 100 repeats of randomly time-permuted data (see text for more details). In this plot, the output singular vector is shifted by -30 to +30 time samples. B. Heatmap of peak correlation coefficients for all of the possible input-output pairs. The singular vectors of input and output belonging to the top 50 matches are ordered differently to maximize the matched pair correlation coefficients. For details of the sorting procedure see text. C. Same as B for randomly permuted data. D. Red: Histogram of correlation coefficients for 152 identified pairs from original data. Blue: Histogram of highest peak correlation coefficient for 100 randomly permuted trials. A Gaussian function can be fitted on this histogram. The threshold is set to be three standard deviation from mean of these 100 points. This leads to the selection of 31 pairs with significant peak correlation coefficient.}
      \label{fig:eigmatch}
  \end{figure*} 
 
    \begin{figure*}
         \centering
         \includegraphics[scale=.25,clip,trim=.5in 3.65in .2in 0in]{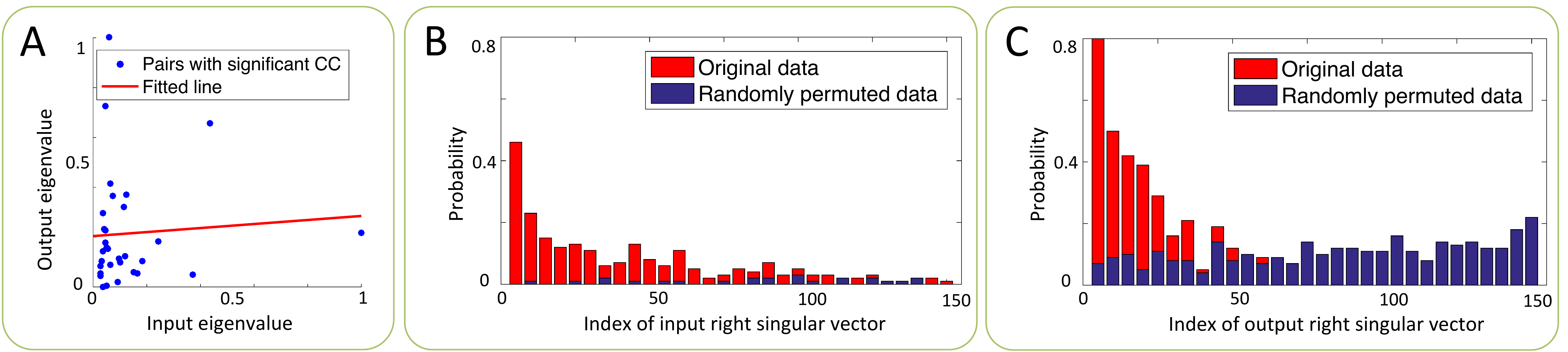}
         \caption{A. Scatterplot of top selected eigenvalues and fitted linear function. The slope of the fitted line is 0.08. B. Probability of selecting each right singular vector by the selection process (see text for more details) for stimulus matrix. Singular vectors are ordered by values of the corresponding singular values (high-to-low). C. Same as B for ganglion cell firing rate matrix.}
         \label{fig:tr}
     \end{figure*} 

Therefore, we came up with the following procedure. We calculate the SVD of ${\bf X}={\bf U}_{X}{\bf \Lambda}_{X}{\bf V}_{X}^T$ and ${\bf Y}={\bf U}_{Y}{\bf \Lambda}_{Y}{\bf V}_{Y}^T$ from data. We match right singular vectors of $\bf X$ and $\bf Y$, columns of ${\bf V}_{ X}$ and ${\bf V}_{ Y}$, using CC as a similarity measure between these vectors. We compute CC for all possible pairs of input-output singular vectors and rank each pair based on this similarity measure. 

To address the delay of neural response to stimulus, CC is computed for several time shifts between stimuli and firing rates and the peak value is considered as the similarity. Fig. \ref{fig:eigmatch}-A (red curve) shows the CC between the pair of singular vectors with the highest peak value. There is a considerable correlation between the two vectors, when compared to peak CC values for randomly permuted data (see also below). 

Finally we use a greedy sorting procedure to identify top pairs. We select the pair with the highest rank as the first pick. This pair corresponds to certain index of input and output singular vectors. We then remove all the other pairs coming from these indexes and repeat a similar process for the remaining pairs. Fig. \ref{fig:eigmatch}-B shows the heatmap of similarity measures for sorted pairs. The diagonal structure implies that there is an approximate one-to-one mapping between the top sorted singular vectors. 

To test the significance of our matching between singular values, we provide a reference by breaking the temporal relationship between stimuli and firing rates and redoing the similarity analysis. We randomly permute columns in $\bf Y$ (neural response matrix) while we keep $\bf X$ (stimulus matrix) the same. This dataset is called the \textit{randomly permuted} data in this paper. We redo the similarity analysis for 100 trials of this random permutation. Fig. \ref{fig:eigmatch}-A shows 100 CC curves with the highest peak for each of these randomly permuted data. The red curve which corresponds to the original data has higher peak CC compared to other curves. Fig. \ref{fig:eigmatch}-C shows the heatmap of peak CC values for one repeat of randomly permuted data. Comparing this figure to fig. \ref{fig:eigmatch}-B, we observe that correlations in the original data are higher compared to the randomly permuted case. 

To identify pairs with significantly higher peak CC in original data, it is necessary to find a threshold CC based on randomly permuted data. Figure \ref{fig:eigmatch}-D shows the process to find this threshold. We first take the highest peak CC from 100 randomly permuted trials. Then, the threshold is set to be three standard deviations from mean of these 100 points. This leads to the selection of 31 pairs with significant peak CC. Out of these 31 input-output pairs, 12 was among the top 31 input singular values and 20 was among the top 31 output singular values. 

To summarize our findings in this section up until now, 1) we found that there is a significant correlation between some pairs of stimulus and firing rate right singular vectors, 2) we can use these correlations to come up with a one-to-one mapping between the singular vectors, and 3) the one-to-one mapping we obtained suggests that the top principal components of the input is represented at the output, consistent with whitening theory.


After identifying the top input-output pairs of singular vectors, which also gives a mapping between covariance matrix eigenvalues, we go back to the question we asked in the previous section: does the power spectrum of retinal ganglion cells decay less steeply that the power spectrum of natural scenes? In Fig. \ref{fig:tr}-A, we show the scatterplot of the normalized covariance eigenvalues with our new matching and the fitted linear function. This should be compared to Figure \ref{fig:psd_decay}-B. We found that the slope is lower in this case, 0.08, suggesting a flattening effect at the output.
%

To test the validity of our results, we perform further analyses. We estimate the probability of a singular vector being selected by our selection process (described above). First,  we divide out the data into 20 non-overlapping segments and repeat the pair identification process for these subsets of data. The probability of selecting each singular value is estimated by dividing the total number of selection for that singular value by 20. To provide a reference, we randomly permute columns in each of the neural response subset matrices while keeping stimulus subset matrix the same and re-estimate the probabilities. Figure \ref{fig:tr}-B and \ref{fig:tr}-C show the probability of selection for top 150 right singular vectors of the stimulus and firing rate matrices, respectively. We find that the top singular vectors, both for stimuli and firing rates, have a significantly higher than chance level probability of selection. These figures suggest that top right singular vectors of the stimulus matrix are highly expected to have a significant CC with the top right singular values of the ganglion cell firing rate matrix. This observation is consistent with the theory that ganglion cell activity represents a projection of visual stimulus onto its principal subspace. 


\section{Conclusion}

In this paper, we put three key predictions of whitening theory to test. We found that 1) retinal ganglion cells outputs are not perfectly uncorrelated (also see\cite{puchalla2005redundancy,pitkow2012decorrelation}), 2) the power spectrum of ganglion cell firing is not completely flat, but rather slopes downwards and 3) there are significant correlations between top right singular vectors of visual stimuli and retinal ganglion cell firing rates, suggesting that retinal output transmits dynamics in the principal subspace of visual scenes. Overall, our findings suggest that whitening alone cannot fully explain response properties of retinal ganglion cells.


\section*{Acknowledgment}

The authors would like to thank Michael Berry, Jason Prentice and Mark Ioffe for providing the dataset and related instructions. Bin Yu acknowledges partial research support from National Science Foundation (NSF) Grant CDS/E-MSS 1228246, and the Center for Science of Information, an NSF Science and Technology Center, under Grant Agreement CCF-0939370.


\end{document}